\documentclass[aps,twocolumn,floats,prd,nofootinbib,superscriptaddress]{revtex4-1}

\usepackage[utf8]{inputenc}

\usepackage{graphicx,amsmath,amsfonts,amssymb,slashed}
\usepackage{bbold,wasysym}

\usepackage{amsmath,amsfonts,bbm,euscript,hyperref}

\usepackage{xcolor}
\usepackage{color}
\usepackage{hyperref}

\hypersetup{colorlinks, citecolor=blue, linkcolor=red, urlcolor=blue}

\usepackage{graphicx,capt-of}

\usepackage{amsmath}
\usepackage{lipsum}

\usepackage[english]{babel}
\usepackage{blindtext}

\newcommand{\bs}{\boldsymbol}

\newcommand\mnras{MNRAS}             % Monthly Notices of the Royal Astronomical Society

\begin{document}

\title{Can Conformally Coupled Modified Gravity Solve The Hubble Tension?}

\author{Tal Abadi}
\affiliation{Department of Physics, Ben-Gurion University of the Negev, Be'er Sheva 84105, Israel}

\author{Ely D. Kovetz}
\affiliation{Department of Physics, Ben-Gurion University of the Negev, Be'er Sheva 84105, Israel}

\begin{abstract} 
The discrepancy between early-Universe inferences and direct measurements of the Hubble constant, known as the Hubble tension, recently became a pressing subject in high precision cosmology. As a result, a large variety of theoretical models have been proposed to relieve this tension. In this work we analyze a conformally-coupled modified gravity (CCMG) model of an evolving gravitational constant due to the coupling of a scalar field to the Ricci scalar, which becomes active around matter-radiation equality, as required for solutions to the Hubble tension based on increasing the sound horizon at recombination. The model is theoretically advantageous as it has only one free parameter in addition to the baseline $\Lambda$CDM ones.
Inspired by similar recent analyses of  so-called early-dark-energy  models,  we constrain the CCMG model using a combination of early and late-Universe cosmological datasets. In addition to the \emph{Planck} 2018 cosmic microwave background (CMB) anisotropies and weak lensing measurements, baryon acoustic oscillations and the Supernova H0 for the Equation of State datasets, we also use large-scale structure (LSS) datasets such as the Dark Energy Survey year 1 and the full-shape power spectrum likelihood from the Baryon Oscillation Spectroscopic Survey, including its recent analysis using effective field theory, to check the effect of the CCMG model on the (milder) $S_8$ tension between the CMB and LSS. We find that the CCMG model can slightly relax the Hubble tension, with $H_{0}=69.6\pm1.6$ km/s/Mpc at 95\% CL, while barely affecting the $S_8$ tension. However, current data does not exhibit strong preference for CCMG over the standard cosmological model. Lastly, we show that the planned CMB-S4 experiment will have the sensitivity required to distinguish between the CCMG model and the more general class of models involving an evolving gravitational constant.
\end{abstract}

\maketitle

%%%%%%%%%%%%%%%%%%%%%%%%%%%%%%%%%%%%%%%%%%%%%%%%%%%%%%%%%%%%%%%%%%%%%%%%%%%%%
%%%%%%%%%%%%%%%%%%%%%%%%%%%%%%%%%%%%%%%%%%%%%%%%%%%%%%%%%%%%%%%%%%%%%%%%%%%%%
\section{Introduction}
%%%%%%%%%%%%%%%%%%%%%%%%%%%%%%%%%%%%%%%%%%%%%%%%%%%%%%%%%%%%%%%%%%%%%%%%%%%%%
%%%%%%%%%%%%%%%%%%%%%%%%%%%%%%%%%%%%%%%%%%%%%%%%%%%%%%%%%%%%%%%%%%%%%%%%%%%%%

The standard $\Lambda$ cold-dark-matter (CDM) cosmological model has been tested by numerous probes and has provided a remarkable explanation for cosmological observations such as the cosmic microwave background (CMB) anisotropies and the baryon acoustic oscillations (BAO). However, despite the immense successes of the $\Lambda$CDM model, there has been a growing discrepancy between the measured values of the Hubble constant $H_0$---the current expansion rate of the Universe---as inferred from early-Universe probes, which assume the $\Lambda$CDM model, and late-Universe probes, which do not assume such a  model.

Most of the late-Universe measurements constrain the value of $H_0$ by applying the distance-ladder method~\cite{Sandage:2006cv}. This method uses parallax measurements to characterize nearby stars (e.g.\ Cepheid-variable stars, ``tip of the red giant branch" (TRGB) stars, Miras, etc.), which are then used to calibrate the luminosity of nearby Type-Ia supernovae (SNe), allowing distant SNeIa to be used to estimate the Hubble flow. The Supernova H0 for the Equation of State (SH0ES) collaboration, which uses Cepheids, recently obtained $H_0=74.03\pm1.42$ km/s/Mpc \cite{Riess:2019cxk}. Other distance-ladder measurements lead to other values, most of them in rough agreement with SH0ES.

The measurement of CMB anisotropies, assuming the $\Lambda$CDM model, allows an indirect inference of the Hubble constant. Inferring the angular size of the sound horizon and constraining the matter and baryon energy densities directly from the CMB temperature, polarization and lensing power spectra, allows the \emph{Planck} 2018 collaboration to determine $H_0=67.36\pm0.54$ km/s/Mpc \cite{Aghanim:2018eyx}. A similar early-Universe approach can be taken using a combination of measurements without including CMB anisotropies: Big Bang nucleosynthesis (BBN); BAO; the FIRAS CMB global temperature and late-Universe measurement (e.g.\ galaxy-lensing based) of the matter density. The result of such analyses agrees quite precisely with that of the CMB~\cite{Verde:2019ivm}. The value of $H_0$ inferred from the CMB measurements is in $4.4\sigma$ tension with the value reported by SH0ES, and a similar discrepancy is present in the majority of the $H_0$ values inferred from other variations of early and late Universe measurements~\cite{Verde:2019ivm}.

Various theoretical solutions were hitherto suggested to solve the $H_0$ discrepancy, which can crudely be divided into two approaches: pre-recombination and post-recombination solutions. A recent review of the Hubble tension \cite{Knox:2019rjx} argued that the pre-recombination solutions are more likely to work, mainly due to the fact that post-recombination solutions affect only the inferred value of $H_0$, while the combined data from BAO and local $H_0$ measurements implies that a reduction of the sound horizon at last scattering is required as well (see, however, Ref.~\cite{Jedamzik:2020zmd}). In particular, it was argued that the critical epoch for achieving such reduction of the sound horizon takes place just prior to recombination.  An increasing number of models  aim to realize such solutions. 

Recent analyses \cite{Hill:2020osr,Ivanov:2020ril,DAmico:2020kxu,Smith:2020rxx} of a popular subclass of these models, referred to as ``early dark energy" (EDE)~\cite{Poulin:2018dzj,Smith:2019ihp,Agrawal:2019lmo,Alexander:2019rsc,Lin:2019qug,Sakstein:2019fmf,Niedermann:2019olb,Kaloper:2019lpl,Berghaus:2019cls}, showed that while they reduce the $H_0$ discrepancy, full cosmological concordance is not restored  due to their tendency to increase the $S_8$ discrepancy between CMB and large-scale-structure  observables, as described below.

In this work we focus on another model suggested to resolve the $H_0$ tension, based on a subclass of scalar-tensor theory. 
This modified gravity (MG) family of models~\cite{Braglia:2020iik,Ballesteros:2020sik,Ballardini:2020iws,Zumalacarregui:2020cjh}, implemented by the coupling of a homogeneous scalar field to the Ricci scalar, acts to increase Newton's gravitational constant $G_N$ prior to matter-radiation equality $z_{eq}$ (that takes place just before recombination), which increases the Hubble parameter (i.e.\ the expansion rate) prior to recombination. The increase in $H(z)$ prior to recombination reduces the sound horizon $r_s$ and increases the inferred value of $H_0$. The scalar field, initially frozen at some initial value, subsequently decays to zero, lowering the value of $G_N$ to its current value during post-recombination era. We emphasize that this ``natural" occurrence is in contrast to what happens in EDE models, where a fine-tuned parameter $z_c$ determines when the EDE component becomes dominant.

The MG model is parameterized by the initial value of the field $\phi_i$ and the coupling constant $\xi$. Together, these parameters determine the deviation in Newton's constant $\Delta G_N\approx-\xi\phi_i^2/M_P^2$, from BBN to present time.
We will focus here on a special case of a conformally-coupled (CC) MG model, for which $\xi\equiv-1/6$ is fixed. The CCMG model thus introduces only one additional parameter, $\phi_i$, compared to $\Lambda$CDM (and two fewer than the popular EDE model).

Although it offers a more natural and simple realization of the solution to the $H_0$ discrepancy, the CCMG model exhibits most of the deficiencies manifested in the EDE models, however to a lower extent. In  EDE models, some of the $\Lambda$CDM parameters shift significantly in order to preserve the fit to the CMB data, while the CCMG model tends to more delicate shifts of these parameters. 

The increase in the Hubble parameter generally acts to slightly suppress the the growth of perturbations, for the modes within the sound horizon, during the period of enhanced expansion. In the EDE scenario this suppression forces a shift upward in $\Omega_ch^2$, so as to compensate for the loss of efficiency in the perturbations growth, while $n_s$ shifts upwards due to the localized contribution of the EDE component to the expansion rate, as detailed in Ref. \cite{Hill:2020osr}. On the other hand, the increase in the gravitational strength, in the CCMG scenario, already acts to compensate for this loss and therefore allows a smaller shift in the value of $\Omega_ch^2$ \cite{Braglia:2020iik,Ballesteros:2020sik}. Furthermore, since the deviation in $G_N$ under the CCMG model is not as localized in  redshift-space as the dominant period of the EDE component, $n_s$ also does not shift as much. Another impact of CCMG, due to the stronger gravitation, is the downwards shift of the matter density $\Omega_{m}$ which reduces significantly, compared to the EDE scenario.

The increases in $\Omega_ch^2$ and $n_s$ increase the late-time amplitude of the density fluctuations $\sigma_8$, aggravating the current (mild) tension between LSS and CMB inferences of this parameter. We follow the convention in Ref. \cite{Joudaki:2019pmv} to quantify the parameter shifts and the associated LSS-CMB tension by the combination of the parameters defined as $S_8 \equiv \sigma_8 (\Omega_m/0.3)^{0.5}$, where $\sigma_8$ is the RMS mass fluctuations in a 8Mpc$/h$ at $z\!=\!0$. LSS experiments  \cite{Abbott:2017wau,Hildebrandt:2016iqg,Hildebrandt:2018yau,Hikage:2018qbn} place a combined constraint of $S_8 = 0.770^{+0.018}_{-0.016}$, which is in about $2.7\sigma$ tension with the \emph{Planck} 2018 CMB result. The results of the analysis of the EDE model in Ref. \cite{Hill:2020osr}, which considered joint CMB-LSS constraints, showed that the EDE model may increase the tension in up to 35\%, compared to the $\Lambda$CDM model. We will see that the effect on $S_8$ of the CCMG model is much weaker.

In this work we consider the constraints on the CCMG model from different data sets composed of CMB, LSS and $H_{0}$ measurements. We find that overall the one-parameter CCMG model exhibits similar properties to those of the three-parameter EDE model, only more moderate. It allows a smaller increase in the value of $H_{0}$ at the cost of much smaller increase in the value of $S_{8}$. The CCMG model is not favorable to $\Lambda$CDM by \emph{Planck} primary-CMB data alone, but the inclusion of CMB lensing + BAO + redshift-space distortions (RSD) + SNIa + SH0ES datasets results in a detectable CCMG component (i.e.\ a non-zero $\phi_i$). The inclusion of DES-Y1 in the joint dataset places stronger constraints on the CCMG parameter, which hints at the difficulty of reconciling LSS and CMB data.
However, when using the EFT-based LSS constraints we find an even more significant CCMG component which results in a better fit to SH0ES data without worsening the fits to CMB and LSS datasets, compared to $\Lambda$CDM, suggesting that EFT-based LSS measurements place weaker constraints compared to the ones from DES-Y1.

We conclude with a simple Fisher analysis to forecast the constraints on the CCMG model from the planned CMB-S4 experiment. In particular we show that it will be able to constrain $\xi\sim-1/6$ to high accuracy, thus distinguishing the CCMG model from other MG models. 

%%%%%%%%%%%%%%%%%%%%%%%%%%%%%%%%%%%%%%%%%%%%%%%%%%%%%%%%%%%%%%%%%%%%%%%%%%%%%
%%%%%%%%%%%%%%%%%%%%%%%%%%%%%%%%%%%%%%%%%%%%%%%%%%%%%%%%%%%%%%%%%%%%%%%%%%%%%
\section{Model}
%%%%%%%%%%%%%%%%%%%%%%%%%%%%%%%%%%%%%%%%%%%%%%%%%%%%%%%%%%%%%%%%%%%%%%%%%%%%%
%%%%%%%%%%%%%%%%%%%%%%%%%%%%%%%%%%%%%%%%%%%%%%%%%%%%%%%%%%%%%%%%%%%%%%%%%%%%%

The inference of $H_0$ from CMB measurement requires the determination of three parameters: the sound horizon $r_s^\star$, the angular diameter distance $D_A^\star$ and the angular acoustic scale $\theta_s^\star$, where ``$^{\star}$" denotes last-scattering. These are related by
%\begin{equation}
$\theta_s^\star\equiv r_s^\star/D_A^\star$,
%\end{equation}
which is measured by \emph{Planck} 2018 to about $0.03\%$ precision \cite{Aghanim:2018eyx}. Thus, any modified evolution of $H(z)$ must accommodate the fixed ratio between $r_s^\star$ and $D_A^\star$.
The sound horizon at last-scattering 
\begin{equation}
r_s^\star=\int_{z_\star}^\infty\frac{dz}{H(z)}c_s(z),
\end{equation}
depends on the evolution of $H(z)$ prior to recombination, while the angular diameter distance
\begin{equation}
D_A^\star=\int_0^{z_\star}\frac{dz}{H(z)}
\end{equation}
depends on its evolution post-recombination (and can be used to set  $H_0$). Writing the expression for $\theta_s^\star$ explicitly:
\begin{equation}
\theta_{s}^{\star}\approx\frac{\intop_{z_{\star}}^{\infty}\frac{dz}{\sqrt{ G_{N}(z)}\sqrt{\rho_{r,0}\left(1+z\right)^{4}+\rho_{m,0}\left(1+z\right)^{3}+\rho_\phi}}c_{s}\left(z\right)}{\intop_{0}^{z_{\star}}\frac{dz}{\sqrt{G_{N}(z)}\sqrt{\rho_{r,0}\left(1+z\right)^{4}+\rho_{m,0}\left(1+z\right)^{3}+\rho_{\Lambda}}}},\label{eqn:theta_s-explicit}
\end{equation}
where we omitted $\rho_{\Lambda}$ and $\rho_{\phi}$ in the higher (numerator)  and lower (denominator) redshift regimes, in which they are negligible, respectively. It is easy to see how an increase in the value of $G_N$ or an introduction of a new dominant contribution to the energy density budget (as EDE models suggest), prior to recombination, acts to reduce the sound horizon while increasing $H(z)$. The CCMG model introduces a scalar field with non-minimal-coupling (NMC), causing an upward shift in the value of Newton's gravitational constant $G_N$ prior to recombination. Around matter-radiation equality the field becomes dynamic and decays, reducing Newton's constant back to its fiducial value. The increase in the gravitational strength enhances the growth of $H(z)$ during this period, which in turn reduces $r_s^\star$ and raises the inferred value of $H_0$. Naively speaking, a deviation of about 15\% in the value of $G_N$ (while keeping all the other parameters fixed and neglecting the additional energy component) is enough to reduce $r_s^\star$ by 7\%, as was suggested in Ref. \cite{Knox:2019rjx} in order to solve the $H_{0}$ tension.

The MG model we consider is described by the action
\begin{equation}
S=\frac{1}{2}\intop d^{4}x\sqrt{-g}\left[\frac{F\left(\phi\right)}{2}R+\partial_{\mu}\phi\partial^{\mu}\phi+\mathcal{L}_{m}\right],\label{eqn:CCMG-action}
\end{equation}
where $F\left(\phi\right)\equiv M_{P}^{2}\left(1+\xi\frac{\phi^{2}}{M_{P}^{2}}\right)$
is the effective Planck mass (i.e. the NMC to the Ricci scalar $R$) and $\mathcal{L}_m$ is the Lagrangian density describing the remaining contents of the Universe. The field $\phi$ is coupled to the Ricci scalar through a dimensionless coupling constant $\xi$, while in the special case of CCMG we fix $\xi=-1/6$, for which the action of the scalar field $\phi$ is invariant under conformal transformations in 4 space-time dimensions and the number of additional parameters to the $\Lambda$CDM model reduces from two ($\xi$ and $\phi_i$) to one. The dynamics of the field $\phi$ are determined by the equation of motion
\begin{equation}
\ddot{\phi}+3H\dot{\phi}-\xi R\phi = 0. \label{eqn:CCMG-EOM}
\end{equation}
We demand that $\xi<0$, therefore as long as $R\ll H $, the field remains frozen at its initial value $\phi_i$.

\begin{figure}
\centering
\includegraphics[width=1\columnwidth]{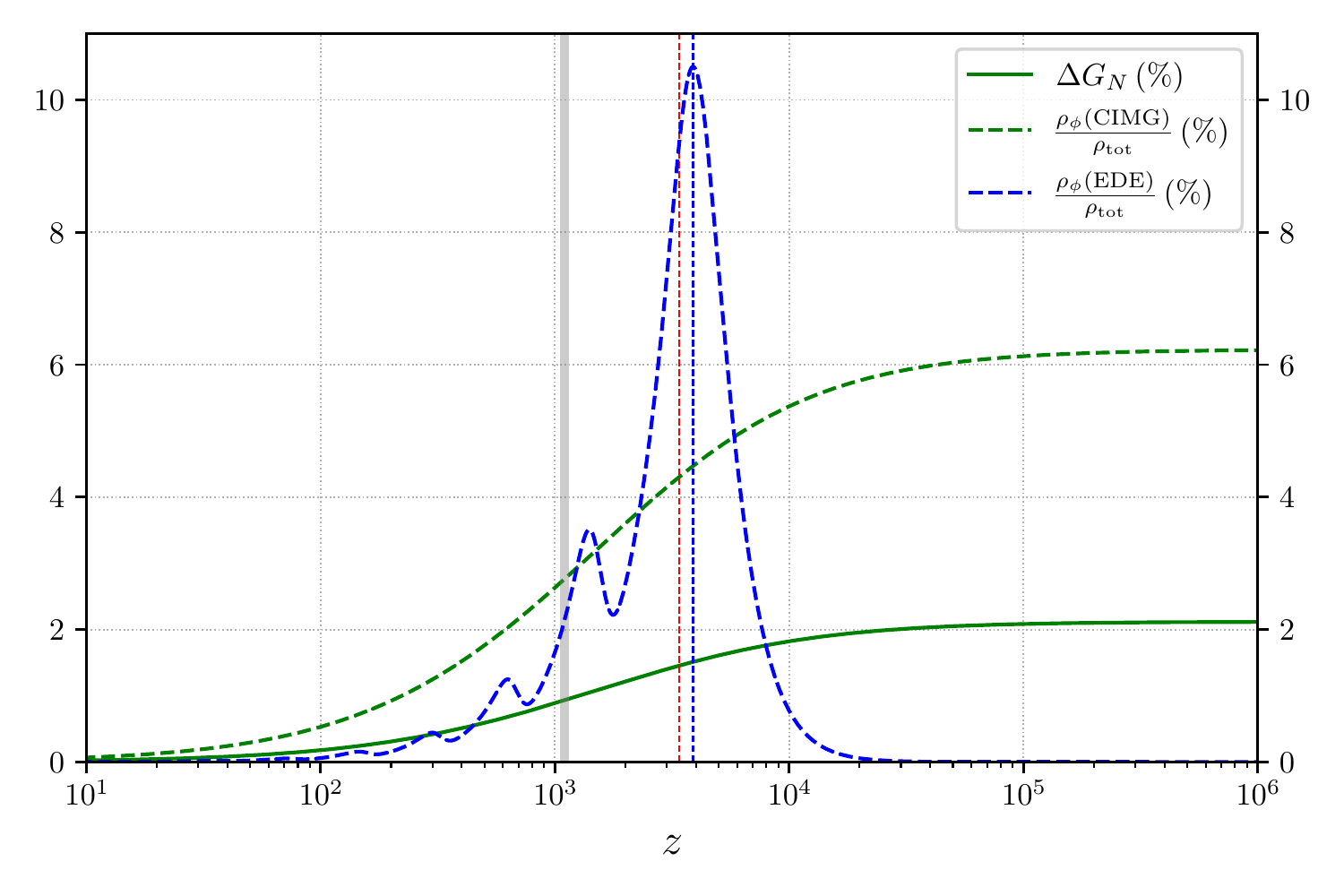}
\caption{Evolution of the relative deviation of Newton's constant $\Delta G_{N}\left(\%\right)=100\times\left|G_{N}^{*}/G_{N}-1\right|$ (solid green) and the energy fraction associated with $\phi$ (dashed green), in the CCMG scenario ($\xi=-1/6$), using the parameters specified in Table~\ref{tab:P18+Lens+EFT+BAO+SH0ES}. For comparison, we also plot the energy fraction associated with the EDE field (dashed blue), corresponding to the best-fit values from Table IV in Ref.~\cite{Hill:2020osr}. We also mark the matter-radiation equality (dashed red), the critical redshift of the EDE model ($z_c$ parameter, vertical dashed blue), in which the EDE field is most dominant, and the recombination epoch (gray band). Both models become dynamic just prior to recombination (around matter-radiation equality). The EDE component is transiently dominant near $z_c$ (peaks to over $10\%$), in contrast to the CCMG component which is present during the entire early-Universe era and is less significant (about $6\%$). The CCMG and EDE models were simulated by a modified version of hi-CLASS and the code of CLASS-EDE, respectively. 
\label{fig:CCMG-background}}
\end{figure}

The evolution of the Ricci scalar can be derived from Einstein's equation
\begin{equation}
R_{\mu\nu} - \frac{1}{2}g_{\mu\nu}R = 8\pi G_N T_{\mu\nu}.\label{eqn:Einstein-eq}
\end{equation}
Since the trace of the stress-energy tensor $T_{\mu\nu}$ vanishes for radiation-like components and is equal to $\rho_m$ for matter-like components, by taking the trace of Eq.~\eqref{eqn:Einstein-eq} we find that $R\propto \rho_m$. Therefore, the Ricci scalar is practically zero---compared to $H^2$---during the radiation-dominated (RD) era. 
Thus the non-minimally coupled field becomes dynamic only around matter-radiation equality when it acquires an effective mass $m^2_\phi\sim\xi R\sim \xi H^2$. Then it begins to roll towards its minimum value, as shown in Fig. \ref{fig:CCMG-background}.
From the Friedmann equation
\begin{equation}
3FH^{2}=\rho+\frac{\dot{\phi}^{2}}{2}+\Lambda-3\dot{F}H\equiv\rho+\rho_{\phi},\label{eqn:MG-general-Friedmann-eq}
\end{equation}
we may associate the extra terms as the energy density of the field $\phi$, so that the energy density of the field,
\begin{equation}
\rho_{\phi}=\frac{1}{2}\dot{\phi}^{2}-6\xi H\phi\dot{\phi}-3\xi H^{2}\phi^{2},\label{eqn:phi-energy-density}
\end{equation}
scales as $a^{-4.5}$ during the matter-dominated (MD) era, and thus dilutes faster than radiation.

Because of the NMC, Newton's constant is replaced by an \emph{effective Newton constant} $G_{N}^{*}\equiv\left(8\pi F\right)^{-1}$, and the deviation from General Relativity (GR) can be parameterized by $\Delta G_N \equiv \left|G_N^*/G_N-1\right|$. The deviation from GR can also be parameterized by means of the so-called Post-Newtonian (PN) parameters \cite{Boisseau:2000pr}
\begin{align}
\gamma_{PN} & =1-\frac{F_{,\phi}^{2}}{F+2F_{,\phi}^{2}},\\
\beta_{PN} & =1+\frac{FF_{,\phi}}{8F+12F_{,\phi}^{2}}\frac{d\gamma_{PN}}{d\phi},\nonumber 
\end{align}
where the prediction from GR, i.e. $\gamma_{PN}=\beta_{PN}=1$, is
tightly constrained by Solar System experiments, as $\gamma_{PN}-1=\left(2.1\pm2.3\right)\times10^{-5}$
and $\beta_{PN}-1=\left(4.1\pm7.8\right)\times10^{-5}$ at 68\% CL \cite{Bertotti:2003rm,Will:2014kxa}.
Recent analysis, Ref.~\cite{Braglia:2020iik}, showed that a deviation of about $2\%$  in $G_N$ at early times is enough to raise the Hubble constant to a value of $H_0=70.56$ km/s/Mpc, without conflicting with Solar System measurements.

%%%%%%%%%%%%%%%%%%%%%%%%%%%%%%%%%%%%%%%%%%%%%%%%%%%%%%%%%%%%%%%%%%%%%%%%%%%%%
%%%%%%%%%%%%%%%%%%%%%%%%%%%%%%%%%%%%%%%%%%%%%%%%%%%%%%%%%%%%%%%%%%%%%%%%%%%%%
\section{Methodology and Datasets}
%%%%%%%%%%%%%%%%%%%%%%%%%%%%%%%%%%%%%%%%%%%%%%%%%%%%%%%%%%%%%%%%%%%%%%%%%%%%%
%%%%%%%%%%%%%%%%%%%%%%%%%%%%%%%%%%%%%%%%%%%%%%%%%%%%%%%%%%%%%%%%%%%%%%%%%%%%%

We implement the CCMG model by modifying the public code for scalar-tensor theories \texttt{hi-class} \cite{Zumalacarregui:2016pph,Bellini:2019syt}\footnote{\url{http://miguelzuma.github.io/hi_class_public/}}, which in turn is based on the public Boltzmann code \texttt{CLASS} \cite{Lesgourgues:2011re,Blas:2011rf}. In particular, we modified the Brans-Dicke model to match the CCMG model by replacing the Brans-Dicke related $G_2$ and $G_4$ (see Ref. \cite{Zumalacarregui:2016pph}) functions with $G_2 = X - \Lambda$ and $G_4 = (1 + \xi\phi^2)/2 $ (and $G_3=G_5=0$), and their corresponding derivatives, where $X\equiv -\partial_\mu\phi\partial^\mu\phi/2$ and $\Lambda$ plays the role of the cosmological constant. 

We also added an extraction of $f\sigma_8(z)$ 
(taken from the public code \texttt{CLASS-EDE} \cite{CLASS-EDE}\footnote{J. C. Hill, E. McDonough and M. W. Toomey: \emph{Class-ede} \url{https://github.com/mwt5345/class_ede}}), 
where $f\equiv d\log D/d\log a$ is the linear growth rate which is needed for implementing the RSD likelihoods in our analyses.
In all likelihoods requiring calculations of the non-linear matter power spectrum, we used the ``Halofit" prescription \cite{Smith:2002dz,Takahashi:2012em} implemented in CLASS.
We followed the analyses in \cite{Hill:2020osr}, performing Markov-Chain-Monte-Carlo (MCMC) analyses, sampling the posterior distributions using the Metropolis-Hasting algorithm \cite{Lewis:2002ah,Lewis:2013hha,neal2005taking}, implemented in the public code \texttt{Cobaya} \cite{Torrado:2020dgo}\footnote{\url{https://cobaya.readthedocs.io/en/latest/}}, with Gelman-Rubin \cite{Gelman:1992zz} convergence criteria $R-1<0.05$. We used a uniform prior for the CCMG parameter $\phi_i=[0.005,1]$, with initial condition $\dot{\phi}_{i}=0$ and Gaussian priors for the $\Lambda$CDM  cosmological parameters, centered around the $\Lambda$CDM fiducial values. For the dataset combination which includes the EFT of LSS, we used MontePython \cite{Audren:2012wb,Brinckmann:2018cvx} for the MCMC analysis, along with the public code PyBird \cite{DAmico:2020kxu}\footnote{\url{https://pybird.readthedocs.io/en/latest/}}, which implements the EFT likelihood. We also used the public code \texttt{GetDist} \cite{Lewis:2019xzd}\footnote{\url{https://getdist.readthedocs.io/en/latest/}} to analyze the MCMC chains: extract the best-fit parameters, mean values and errors, and plot the correlations in parameter-space as well as the maximized posteriors.

We used the same datasets used in Ref.~\cite{Hill:2020osr}, which include: \emph{Planck} 2018 CMB temperature and polarization anisotropies power spectra (TT + TE + EE) and the CMB lensing (P18 and lensing) \cite{Aghanim:2018eyx}, Baryonic Acoustic Oscillations (BAO) \cite{Alam:2016hwk,Ross:2014qpa,Beutler:2011hx,Kaiser:1987qv}, redshift-space distortion from SDSS BOSS DR12 (RSD) \cite{Alam:2016hwk,Satpathy:2016tct}, Type-Ia Supernovae (SNIa Pantheon) \cite{Scolnic:2017caz}, SH0ES 2019 $H_{0}$ measurements\footnote{There is a more recent measurement of $H_{0}$ by SH0ES, $H_{0}=73.5\pm1.4$ km/s/Mpc \cite{Reid:2019tiq}, while we used the previously reported value of $H_{0}=74.03\pm1.42$ km/s/Mpc, in order to facilitate a direct comparison to other studies. In any case this difference in the SH0ES $H_0$ does not alter the conclusions of our analysis.} \cite{Riess:2019cxk} and the Dark Energy Survey Year 1 (DES Y1) \cite{Abbott:2017wau,2018MNRAS.480.3879A}, which has not been used until now in an analysis of the CCMG model, with the exception of additional LSS data from the Kilo-Degree Survey \cite{Hildebrandt:2018yau,Hildebrandt:2016iqg} (KiDS) and the Hyper Suprime-Cam \cite{Hikage:2018qbn} (HSC) survey, which we used only as reference. Note that the DES-Y1 likelihoods should in principle be adapted according to the modifications to gravity, however, in the model we analyze, such modifications are expected to be very small, thus we expect their use to be safe.
We also included an additional dataset, comprising of the effective field theory (EFT) of LSS \cite{Perko:2016puo,DAmico:2019fhj,Colas:2019ret,DAmico:2020kxu} applied to the full shape power spectrum of the BOSS/SDSS galaxies clustering DR12 \cite{Gil-Marin:2015sqa,Gil-Marin:2015nqa} and the BAO post-reconstruction measurements from BOSS, combined with covariance between EFT-BOSS and anisotropic BAO analysis. We did not use the south galactic cap (SGC) field of LOWZ, as in Ref.~\cite{DAmico:2019fhj}. This dataset is henceforth referred to as BAO+EFT (not to be confused with the BAO dataset without EFT).

%%%%%%%%%%%%%%%%%%%%%%%%%%%%%%%%%%%%%%%%%%%%%%%%%%%%%%%%%%%%%%%%%%%%%%%%%%%%%
%%%%%%%%%%%%%%%%%%%%%%%%%%%%%%%%%%%%%%%%%%%%%%%%%%%%%%%%%%%%%%%%%%%%%%%%%%%%%
\section{Results}
%%%%%%%%%%%%%%%%%%%%%%%%%%%%%%%%%%%%%%%%%%%%%%%%%%%%%%%%%%%%%%%%%%%%%%%%%%%%%
%%%%%%%%%%%%%%%%%%%%%%%%%%%%%%%%%%%%%%%%%%%%%%%%%%%%%%%%%%%%%%%%%%%%%%%%%%%%%

We adopt the \emph{Planck} convention, holding the sum of  neutrino masses fixed to $0.06\,{\rm eV}$, with one massive eigenstate against two massless eigenstates, and we fix the effective number of relativistic species to $N_{\rm eff}=3.046$. We also fit the $\Lambda$CDM model to each data set as a benchmark for comparison. A summary of the results is tabulated in Table \ref{tab:result-summary}.

%%%%%%%%%%%%%%%%%%%%%%%%%%%%%%%%%%%%%%%%%%%%%%%%%%%%%%%%%%%%%%%%%%%%%%%%%%%%%
\subsection{CCMG Meets Primary CMB Alone}
\label{sec:P18}
%%%%%%%%%%%%%%%%%%%%%%%%%%%%%%%%%%%%%%%%%%%%%%%%%%%%%%%%%%%%%%%%%%%%%%%%%%%%%

\begin{table*}[!htb]
\centering
Constraints on CCMG from \emph{Planck} 2018 data only: TT+TE+EE\\
\vspace{2pt}
\begin{tabular}{|l|c|c|c|}
\hline \hline
Parameter 						& $\Lambda$CDM 			& CCMG 	 \tabularnewline
\hline \hline
$\bs{\log(10^{10}A_{\rm s})}$ 			& $3.044\,(3.046)\pm0.015$ 		& $3.047\,(3.035)\pm0.016$  
\tabularnewline
$\bs{n_{\rm s}}$ 					& $0.9645\,(0.9618)\pm0.0042$ 	& $0.9667\,(0.9638)_{-0.0053}^{+0.0045}$ 
\tabularnewline
$\bs{100\theta_{\rm s}}$				& $1.04185\,(1.0419)\pm0.00029$ 	& $1.04189\,(1.04151)\pm0.00030$ 
\tabularnewline
$100\times\bs{\Omega_{\rm b}h^{2}}$ 	& $2.235\,(2.23284)\pm0.014$ 		& $2.238\,(2.23162)\pm0.015$
\tabularnewline
$\bs{\Omega_{\rm c}h^{2}}$ 			& $0.1202\,(0.1210)\pm0.0013$ 	& $0.1200\,(0.1206)\pm0.0014$
\tabularnewline
$\bs{\tau_{\mathrm{reio}}}$ 			& $0.0540\,(0.0547)\pm0.0075$ 	& $0.0547\,(0.0489)\pm0.0077$
\tabularnewline
$\bs{\phi_{i}}\,[M_P]$ 				& -- 							& $<0.213\, (0.087)$	\\
\hline
$H_{0}\, [\mathrm{km/s/Mpc}]$			& $67.30\,(66.98)\pm0.58$ 		& $67.98\,(67.11)_{-1.1}^{+0.63}$ 
\tabularnewline
$\Omega_{\rm m}$					& $0.3162\,(0.3210)\pm0.0081$ 	& $0.310\,(0.319)_{-0.0091}^{+0.012}$ 
\tabularnewline
$\sigma_{8}$ 						& $0.8112\,(0.8142)\pm0.0072$ 	& $0.8170\,(0.8094)_{-0.020}^{+0.022}$ 
\tabularnewline 
$S_{8}$ 							& $0.833\,(0.842)\pm0.016$	 	& $0.830\,(0.834)\pm0.016$ 
\tabularnewline 
$\Delta G_{N}\,(\%)$					& -- 							& $0.68\,(0.13)_{-0.75}^{+0.14}$
\tabularnewline
\hline 
\end{tabular}
\medskip{}
\caption{The mean (best-fit) $\pm1\sigma$ (68\% CL) constraints on the cosmological parameters in $\Lambda$CDM and CCMG as inferred from \emph{Planck} 2018 primary CMB (TT+TE+EE) data alone. The CCMG component is not significant when considering
 early-Universe information (gravitational lensing have a little influence as well).
\label{tab:CCMG-P18}}
\end{table*}

The first analysis we performed includes only the temperature and polarization anisotropies power-spectrum data from \emph{Planck} 2018. While there is a small contribution to this dataset from LSS, due to gravitational lensing of the power spectra, the overall constraints are dominated by information from the recombination epoch. This analysis tests for evidence for the CCMG model using early-Universe data alone.

The results in Table~\ref{tab:CCMG-P18} (see also Table~\ref{tab:result-summary}), which are consistent with Ref. \cite{Ballardini:2020iws} (see Table 2), show very weak evidence for the CCMG model. The single CCMG parameter is constrained by an upper bound $\phi_i<0.213$ $M_{P}$ (also indicated by the posterior shown in Fig. \ref{fig:CCMG-summary}) and $\Delta G_N$ comprises 0 within $1\sigma$. It seems that primary CMB data alone does not prefer the CCMG model over the standard $\Lambda$CDM model, and indeed the shift in the cosmological parameters is negligible (below $1\sigma$ of the $\Lambda$CDM benchmark).
We also note that introducing an additional parameter beyond  $\Lambda$CDM to the fit in the CCMG parameter model does not improve the fit as one might expect. On the contrary, we find $\Delta\chi^2=+4.8$, as shown in Table \ref{table:chi2_CMB_alone}.
We conclude that there is no preferred region, compared to the $\Lambda$CDM model, within the CCMG parameter space when considering the primary CMB data alone.

\begin{table}[!htb]
\centering
%\scalebox{0.9}{
$\chi^2$ statistics from \emph{Planck} 2018 data only: TT+TE+EE\\
\vspace{2pt}
  \begin{tabular}{|l|c|c|}
    \hline\hline
    			Datasets 						&	$\Lambda$CDM 	&	CCMG		\\
    	\hline\hline
         \textit{Planck} 2018 low-$\ell$ TT			& 	24.1				& 	23.4	\\
        \textit{Planck} 2018 low-$\ell$ EE			&	396.3			& 	395.7 \\
    \textit{Planck} 2018 high-$\ell$ TT+TE+EE		&	2345.1			& 	2351.2	\\

    \hline
    			Total $\chi^2 $   				& 	2765.5			& 	2770.3	\\
    \hline
  \end{tabular}
\medskip{}
  \caption{$\chi^2$ values for the best-fit $\Lambda$CDM and CCMG models, constrained by  primary CMB alone. The additional parameter of the CCMG model does not improve the fit to the data as may be expected when increasing the total number of parameters.  \label{table:chi2_CMB_alone}}

\end{table}

%%%%%%%%%%%%%%%%%%%%%%%%%%%%%%%%%%%%%%%%%%%%%%%%%%%%%%%%%%%%%%%%%%%%%%%%%%%%%
\subsection{Expanding the analysis to also include CMB lensing, BAO, RSD, SNIa, and SH0ES}
%%%%%%%%%%%%%%%%%%%%%%%%%%%%%%%%%%%%%%%%%%%%%%%%%%%%%%%%%%%%%%%%%%%%%%%%%%%%%

\begin{table*}[!htb]
\vspace{0.5cm}
\centering
Constraints from \emph{Planck} 2018 TT+TE+EE + CMB Lensing,\\
BAO, RSD, SNIa and SH0ES\\
\vspace{2pt}
\begin{tabular}{|l|c|c|c|}
\hline \hline
Parameter 						& $\Lambda$CDM 						& CCMG 	 \tabularnewline
\hline \hline
$\bs{\log(10^{10}A_{\rm s})}$ 			& $3.051\,(3.054)_{-0.015}^{+0.013}$ 		& $3.051\,(3.060)\pm0.014$  
\tabularnewline
$\bs{n_{\rm s}}$ 					& $0.9689\,(0.9691)\pm0.0035$ 			& $0.9711\,(0.9727)\pm 0.0038$ 
\tabularnewline
$\bs{100\theta_{\rm s}}$				& $1.04204\,(1.04187)\pm0.00027$ 			& $1.04199\,(1.04212)\pm0.00028$ 
\tabularnewline
$100\times\bs{\Omega_{\rm b}h^{2}}$ 	& $2.253\,(2.253)\pm0.013$ 				& $2.247\,(2.24705)\pm0.013$
\tabularnewline
$\bs{\Omega_{\rm c}h^{2}}$ 			& $0.1183\,(0.1185)\pm0.0009$ 			& $0.1193\,(0.1192)^{+0.0010}_{-0.0011}$
\tabularnewline
$\bs{\tau_{\mathrm{reio}}}$ 			& $0.0593\,(0.0618)^{+0.0065}_{-0.0075}$ 	& $0.0565\,(0.0601)\pm0.0073$
\tabularnewline
$\bs{\phi_{i}}\,[M_P]$ 				& -- 									& $0.297\,(0.297)^{+0.11}_{-0.075}$	\\
\hline
$H_{0}\, [\mathrm{km/s/Mpc}]$			& $68.17\,(68.07)\pm0.39$ 				& $69.24\,(69.15)_{-0.83}^{+0.60}$ 
\tabularnewline
$\Omega_{\rm m}$					& $0.3045\,(0.3057)\pm0.0051$ 			& $0.297\,(0.298)^{+0.007}_{-0.006}$ 
\tabularnewline
$\sigma_{8}$ 						& $0.8088\,(0.8103)\pm0.0059$ 			& $0.8242\,(0.8267)_{-0.012}^{+0.009}$ 
\tabularnewline 
$S_{8}$ 							& $0.815\,(0.818)\pm0.010$	 			& $0.820\,(0.823)\pm0.010$ 
\tabularnewline 
$\Delta G_{N}\,(\%)$					& -- 									& $1.70\,(1.49)_{-1.1}^{+0.81}$
\tabularnewline
\hline 
\end{tabular}
\medskip{}
\caption{The mean (best-fit) $\pm1\sigma$ (68\% CL) constraints on the cosmological parameters in the $\Lambda$CDM and CCMG scenarios, as inferred from the combination of P18 + lensing + BAO + SNIa + RSD + SH0ES datasets. There is significant evidence for the CCMG model as $\phi_{i}=0.3^{+0.14}_{-0.24}$ $M_{P}$ and $\Delta G_{N}(\%)=1.7\pm1.7$ with 95\% CL are detected at $\ge2\sigma$ significance.
\label{tab:P18+Lens+BAO+SN+H0}}
\end{table*}

\begin{table}[!htb]
\centering
%\scalebox{0.9}{
$\chi^2$ statistics from the fit to \emph{Planck} 2018 TT+TE+EE +\\
CMB Lensing, BAO, RSD, SNIa and SH0ES\\
\vspace{2pt}
  \begin{tabular}{|l|c|c|}
    \hline\hline
    			Datasets 									&	LCDM 		&	CCMG	\\
    			\hline \hline
    CMB TT, EE, TE: & &\\
      \;\;\;\; \textit{Planck} 2018 low-$\ell$ TT 					& 22.9			& 22.4		\\
       \;\;\;\; \textit{Planck} 2018 low-$\ell$ EE 					& 398.0 			& 397.3		\\
  \hspace{.16cm} \; \begin{tabular}[t]{@{}c@{}}
  					\textit{Planck} 2018 high-$\ell$ \\ 
  					TT+TE+EE\end{tabular}   				& 2350.9 			& 2347.2		\\

    LSS:& &\\
    \;\;\;\;\,\textit{Planck} CMB lensing							& 8.7 			& 9.1			\\
   \;\;\;\; BAO (6dF)										& 0.002	 		& 0.035		\\
   \;\;\;\; BAO (DR7 MGS)									& 1.6 			& 2.3			\\
      \;\;\;\; BAO+RSD (DR12 BOSS)							& 6.0 			& 6.7			\\
    Supernovae:& &\\
   \;\;\;\; Pantheon 										& 1034.8 			& 1034.7		\\
   SH0ES 												& 20.2  			& 14.4		\\
    \hline
    Total $\chi^2$   										& 3843.1 			& 3834.1 		\\
    \hline
  \end{tabular}
\medskip{}
  \caption{$\chi^2$ values for the best-fit $\Lambda$CDM and CCMG models, constrained by P18 + lensing + BAO + RSD + SNIa + SH0ES datasets. There is reduction of 9 in the value of $\chi^2$, for the one additional CCMG parameter to $\Lambda$CDM, driven almost entirely by the improved fit to SH0ES. However, it is notable that the fit to the LSS data is worse in the CCMG scenario, while the fit to the CMB is not degraded.
\label{tab:chi2_P18+Lens+BAO+SN+H0}}
\end{table}

Following previous analyses \cite{Smith:2019ihp,Hill:2020osr} of the EDE model, we include \emph{Planck} 2018 CMB lensing, BAO, RSD, supernova, and local distance-ladder data in SH0ES 2019. This data set is considered a conclusive combination of early-Universe, LSS and SNeIa distance-ladder data.

We now find significant evidence for the CCMG model, with this  combination of datasets. The initial value of the field, $\phi_{i}=0.3^{+0.14}_{-0.24}$ $M_{P}$ at 95\% CL , and the fractional deviation of Newton's constant, $\Delta G_{N}(\%)=1.7\pm 1.7$, both at 95\% CL, are detected at $\ge2\sigma$. As a result, the value of $H_{0}$ increases to $H_{0}=69.24\pm1.4$ km/s/Mpc, compared to the $\Lambda$CDM benchmark, $H_{0}=68.17\pm0.77$ km/s/Mpc at 95\% CL. The LSS likelihoods of RSD and BAO have large enough error bars to overlap with the region in parameter space with larger value of $H_{0}$, reducing the Hubble tension to $3.1\sigma$. This is due to both the increase in $H_{0}$ and its increased errors, emphasising the difficulty of reconciling all the likelihoods in the dataset.

\begin{figure*}[!htb]
\begin{centering}
\includegraphics[width=2\columnwidth]{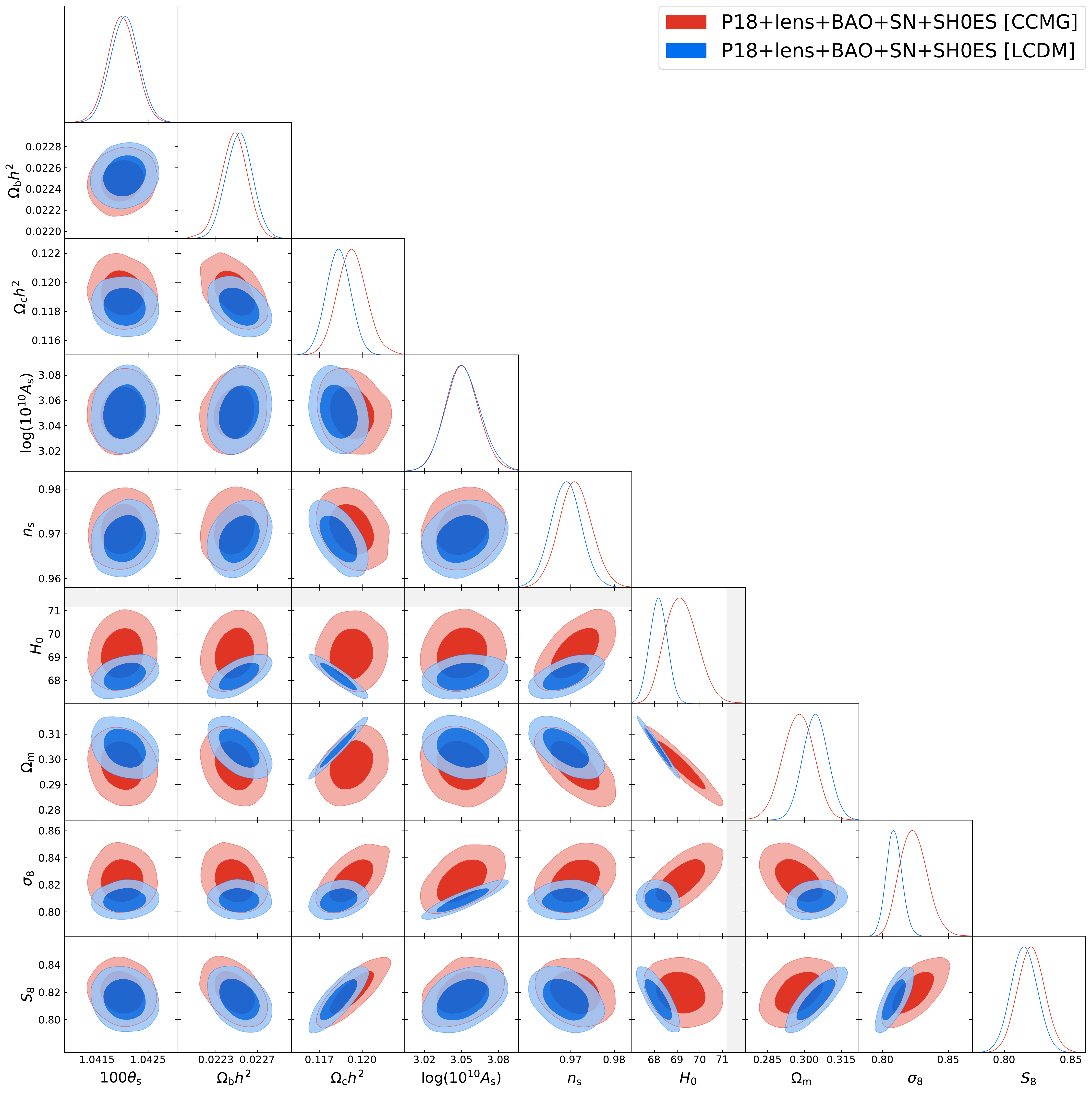}
\par\end{centering}
\caption{Cosmological parameter constraints from the combination of \emph{Planck} 2018 primary CMB data (TT+TE+EE); Planck
2018 CMB lensing data; BAO data from 6dF, SDSS DR7, and SDSS DR12; Pantheon SNIa data; the latest SH0ES $H_0$ constraint; and SDSS DR12 RSD data. We do not plot $\tau$, as it is essentially unchanged in the CCMG fit. Some parameters shift by a non-negligible amount in the CCMG fit (compared to $\Lambda$CDM), including increases in $\Omega_ch^2$, $n_s$, and $\sigma_8$ as well as broadening of the error bars on these parameters. The increase in $H_0$ is not large enough to reconcile the tension with the SH0ES-only constraint (shown in the grey bands), but it does reduce the tension significantly.
\label{fig:P18+Lens+BAO+SN+H0}}
\bigskip
\end{figure*}

In order to keep the fit with CMB and LSS data, other cosmological parameters shift as well. In particular $\Omega_{c}h^{2}$ and $n_{s}$ shift upwards slightly, a $0.7\sigma$ and $0.4\sigma$ discrepancy with the benchmark, respectively. The degeneracy between $H_{0}$ and $\Omega_{c}h^{2}$ breaks due to the introduction of a new energy density component of the CCMG field, while its degeneracy with $n_{s}$ increases (see Fig.~\ref{fig:P18+Lens+BAO+SN+H0}). We also note a minuscule downward shift in $\Omega_{b}h^{2}$, $0.3\sigma$ discrepancy with the benchmark. In addition we find an increase in the value of $\sigma_{8}$ and a decrease in $\Omega_{m}$, for which the net result is a minor increase in $S_{8}$, which translates to a slightly larger tension with the combined LSS constraint: $2.4\sigma$, compared to $2.2\sigma$ for the $\Lambda$CDM benchmark. We find that the correlations between the cosmological parameters and the best-fit $\pm1\sigma$ values, found for this dataset, are in agreement with previous analyses (see Table 2 in Ref. \cite{Ballardini:2020iws} and Figure 5 in Ref. \cite{Ballesteros:2020sik}).
The CCMG component acts to increase the early-Universe expansion rate, thus suppressing the matter power spectrum $P(k)$ for modes smaller than the sound horizon. This suppression requires an upward shift in $\Omega_ch^2$,  which is the driver of the changes in $P(k)$, translating in a larger $\sigma_8$. Moreover, as the enhanced expansion of the Universe is localized in time, the shift in the matter power spectra is scale dependent, affecting the value of $n_{s}$. Such behavior is expected, to some extent, for every model that acts to increase $H_{0}$ in such manner.

Compared to a recent analysis of the EDE model in Ref. \cite{Hill:2020osr}, the CCMG model exhibits smaller shifts of the cosmological parameters (including $H_{0}$). Since the CCMG component is less localized in time, its free parameter $\phi_{i}$ is less correlated with $n_{s}$ than the corresponding parameters in the EDE scenario. In addition, the enhanced gravitational strength acts to boost density anisotropies, which counteracts the need for increasing $\Omega_c h^2$. As a result, the matter density $\Omega_{m}$ is reduced, due to the increased Hubble parameter and the almost unchanged value of the CDM density. That is in contrast to the EDE model, which exhibits a significant increase in $\Omega_{c}h^{2}$ and no shift downwards in $\Omega_{m}$, which results in a higher $S_{8}$. Therefore the CCMG model offers to relax the $H_{0}$ tension, although not as much as the EDE model, but  almost without worsening the CMB-LSS tension, when quantified in terms of the well-constrained $S_{8}$ parameter.

We find that the additional CCMG parameter improves the total fit to the data, with $\Delta\chi^{2}=-9$, relative to the $\Lambda$CDM benchmark, as shown in Table \ref{tab:chi2_P18+Lens+BAO+SN+H0}. The reduction in $\chi^{2}$ is mainly due to the better fit to the SH0ES likelihood which compensates for the degraded fit to the CMB datasets, while the fit to LSS data worsens only slightly, indicating the intrinsic tension between the datasets.

The different shifts in $H_0$ and $S_8$ values indicate stronger correlation of the EDE component $f_{\rm EDE}$ with $H_0$ and $\sigma_8$ than that of the CCMG parameter $\phi_i$ shown in Fig. \ref{fig:CCMG-summary}. Placing the CCMG model somewhere between $\Lambda$CDM and EDE in context of both $H_0$ and $S_8$ tensions.

%%%%%%%%%%%%%%%%%%%%%%%%%%%%%%%%%%%%%%%%%%%%%%%%%%%%%%%%%%%%%%%%%%%%%%%%%%%%%
\subsection{Considering additional LSS data}
%%%%%%%%%%%%%%%%%%%%%%%%%%%%%%%%%%%%%%%%%%%%%%%%%%%%%%%%%%%%%%%%%%%%%%%%%%%%%

\begin{table*}[!htb]
\vspace{0.5cm}
\centering
Constraints from \emph{Planck} 2018 TT+TE+EE + CMB Lensing,\\
BAO, RSD, SNIa, SH0ES and DES-Y1\\
\vspace{2pt}
\begin{tabular}{|l|c|c|c|}
\hline \hline
Parameter 						& $\Lambda$CDM 						& CCMG 	 \tabularnewline
\hline \hline
$\bs{\log(10^{10}A_{\rm s})}$ 			& $3.049\,(3.043)\pm0.012$		 		& $3.048\,(3.053)\pm0.014$  
\tabularnewline
$\bs{n_{\rm s}}$ 					& $0.9705\,(0.9701)\pm0.0030$ 			& $0.9722\,(0.9728)\pm 0.0037$ 
\tabularnewline
$\bs{100\theta_{\rm s}}$				& $1.04208\,(1.04183)\pm0.00023$ 			& $1.04205\,(1.04179)\pm0.00028$ 
\tabularnewline
$100\times\bs{\Omega_{\rm b}h^{2}}$ 	& $2.259\,(2.258)\pm0.011$ 				& $2.255\,(2.250)\pm0.013$
\tabularnewline
$\bs{\Omega_{\rm c}h^{2}}$ 			& $0.1176\,(0.1179)\pm0.0007$ 			& $0.1183\,(0.1184)\pm0.0009$
\tabularnewline
$\bs{\tau_{\mathrm{reio}}}$ 			& $0.0591\,(0.0544)\pm0.0060$		 	& $0.0569\,(0.0563)\pm0.0072$
\tabularnewline
$\bs{\phi_{i}}\,[M_P]$ 				& -- 									& $0.264\,(0.308)^{+0.13}_{-0.072}$	\\
\hline
$H_{0}\, [\mathrm{km/s/Mpc}]$			& $68.52\,(68.30)\pm0.30$ 				& $69.40\,(69.43)_{-0.75}^{+0.60}$ 
\tabularnewline
$\Omega_{\rm m}$					& $0.2999\,(0.3025)\pm0.0039$ 			& $0.2939\,(0.2937)\pm0.0058$ 
\tabularnewline
$\sigma_{8}$ 						& $0.8056\,(0.8038)\pm0.0047$ 			& $0.8176\,(0.8216)_{-0.0110}^{+0.0087}$ 
\tabularnewline 
$S_{8}$ 							& $0.805\,(0.807)\pm0.007$	 			& $0.809\,(0.813)\pm0.009$ 
\tabularnewline 
$\Delta G_{N}\,(\%)$					& -- 									& $1.36\,(1.61)_{-1.1}^{+0.68}$
\tabularnewline
\hline 
\end{tabular}
\medskip{}
\caption{The mean $\pm1\sigma$ (68\% CL) constraints on the cosmological parameters in $\Lambda$CDM and CCMG, as inferred from the combination of P18 + lensing + BAO + RSD + SNIa + SH0ES + DES-Y1. With the inclusion of DES data the evidence for CCMG decreases, as $\phi_{i}=0.26^{+0.17}_{-0.23}$ $M_{P}$ and $\Delta G_{N}(\%)=1.4_{-1.5}^{+1.7}$ with 95\% CL, to $\lesssim 2\sigma$ significance.
\label{tab:P18+Lens+BAO+SN+H0+DES_Y1}}
\end{table*}

\begin{table}[!htb]
\centering
%\scalebox{0.9}{
$\chi^2$ statistics from the fit to \emph{Planck} 2018 TT+TE+EE +\\
CMB Lensing, BAO, RSD, SNIa, SH0ES and DES-Y1\\
\vspace{2pt}
  \begin{tabular}{|l|c|c|}
    \hline\hline
    			Datasets 									&	LCDM 		&	CCMG	\\
    			\hline \hline
    CMB TT, EE, TE: & &\\
      \;\;\;\; \textit{Planck} 2018 low-$\ell$ TT 					& 22.4			& 22.3		\\
       \;\;\;\; \textit{Planck} 2018 low-$\ell$ EE 					& 396.0 			& 396.3		\\
  \hspace{.16cm} \; \begin{tabular}[t]{@{}c@{}}
  					\textit{Planck} 2018 high-$\ell$ \\ 
  					TT+TE+EE\end{tabular}   				& 2350.6 			& 2350.5		\\

    LSS:& &\\
   \;\;\;\;\,\textit{Planck} CMB lensing							& 9.4 			& 9.2			\\
   \;\;\;\; BAO (6dF)										& 0.002	 		& 0.084		\\
   \;\;\;\; BAO (DR7 MGS)									& 1.9 			& 2.7			\\
   \;\;\;\; BAO+RSD (DR12 BOSS)							& 5.8 			& 7.3			\\
   \;\;\;\; DES-Y1											& 510.8 			& 513.6		\\
    Supernovae:& &\\
   \;\;\;\; Pantheon 										& 1034.8 			& 1034.8		\\
   SH0ES 												& 18.8  			& 13.0		\\
    \hline
    Total $\chi^2$   										& 4350.5 			& 4349.7 		\\
    \hline
  \end{tabular}
\medskip{}
  \caption{$\chi^2$ values for the best-fit $\Lambda$CDM and CCMG models, constrained by CMB + lensing + BAO + RSD + SNIa + SH0ES + DES-Y1 datasets. There is reduction of only 0.8 in $\chi^2$, for the one additional parameter of the CCMG model.
\label{tab:chi2_P18+Lens+BAO+SN+H0+DES_Y1}}
\end{table}

We now expand our analysis to include likelihoods from the DES-Y1 dataset~\cite{Abbott:2017wau,2018MNRAS.480.3879A}, in particular the ``3x2pt" likelihood, comprised of photometric galaxy clustering, galaxy-galaxy lensing, and cosmic shear two-point correlation functions.

The inclusion of DES data places stronger constraints on $\Omega_{m}$, which in turn reduces the value of the CCMG parameter $\phi_{i}$, as shown in Table~\ref{tab:P18+Lens+BAO+SN+H0+DES_Y1} (see also Table~\ref{tab:result-summary}). We find $\phi_{i}=0.26^{+0.17}_{-0.23}$ $M_{P}$ at 95\% CL, detected with $\le2\sigma$ significance. Meanwhile, the value of $H_{0}$ shifts further upwards to $H_{0}\!=\!69.4_{-1.2}^{+1.3}$ km/s/Mpc at 95\% CL. The reason for that is due to the general shift in $H_{0}$ when including the DES-Y1 dataset, observed also for the $\Lambda$CDM benchmark compared to Table~\ref{tab:P18+Lens+BAO+SN+H0}. Thus the tension with SH0ES measurements is reduced to $3\sigma$ for CCMG, compared to $3.8\sigma$ in the $\Lambda$CDM benchmark scenario.

The lower value of $\phi_i$ when DES-Y1 data is included in the combined dataset can be understood in terms of the interplay between $\sigma_8$, $\Omega_m$, $H_0$ and $\phi_i$. The precise DES measurement of $\Omega_m$ breaks the $\Omega_m-H_0$ degeneracy in the $\Lambda$CDM fit to the CMB, shifting $H_0$ to larger values. The impact of the DES measurements on the CCMG parameter results in a lower value for $\phi_i$, due to a marked correlation between $\sigma_8$, $H_0$ and $\phi_i$, observed in Fig.~\ref{fig:CCMG-summary}. The same thing happens in the EDE scenario with $f_{\rm EDE}$ replacing $\phi_i$, only to greater extent due to its stronger degeneracy with $\sigma_8$ and $H_0$. Therefore, the CCMG is less conflicting with DES-Y1 likelihoods than the EDE model.

It is also notable that the posterior of $\sigma_8$ matches closely  that of the fit to primary CMB-only, as shown in Fig. \ref{fig:CCMG-summary}, erasing the shift observed without DES. This shift manifests the constraints of LSS on $\phi_i$, due to the correlation between $\sigma_8$ and $\phi_i$, mentioned previously. The shift in $\sigma_8$ is matched by the shift in $S_{8}=0.809\pm0.018$ at 95\% CL, which is in $1.9\sigma$ tension with combined LSS measurements, negligibly larger than the $1.8\sigma$ for $\Lambda$CDM.

The $\chi^2$ statistics, tabulated in Table \ref{tab:chi2_P18+Lens+BAO+SN+H0+DES_Y1}, show poor improvement to the fit for the additional CCMG parameter. The CCMG model offers a slightly better fit to SH0ES data alone, compared to the $\Lambda$CDM benchmark, while the fit to LSS data worsens. It seems that there is no region in parameter space that is in concordance with all cosmological data sets. This indicates a possible statistical tension between LSS and $H_0$ likelihoods, as each dataset pulls  the parameters in the opposite direction.

\begin{table*}[!htb]
\vspace{0.5cm}
\centering
Constraints from \emph{Planck} 2018 TT+TE+EE + CMB Lensing,\\
BAO + EFT and SH0ES\\
\vspace{2pt}
\begin{tabular}{|l|c|c|c|}
\hline \hline
Parameter 						& $\Lambda$CDM 						& CCMG 	 \tabularnewline
\hline \hline
$\bs{\log(10^{10}A_{\rm s})}$ 			& $3.051\,(3.040)^{+0.014}_{-0.016}$		 & $3.052\,(3.059)\pm0.015$  
\tabularnewline
$\bs{n_{\rm s}}$ 					& $0.9690\,(0.9683)\pm0.0037$ 			& $0.9721\,(0.9751)\pm 0.0041$ 
\tabularnewline
$\bs{100\theta_{\rm s}}$				& $1.04205\,(1.04195)\pm0.00028$ 			& $1.04204\,(1.04189)\pm0.00028$ 
\tabularnewline
$100\times\bs{\Omega_{\rm b}h^{2}}$ 	& $2.252\,(2.260)\pm0.013$ 				& $2.249\,(2.247)\pm0.013$
\tabularnewline
$\bs{\Omega_{\rm c}h^{2}}$ 			& $0.1182\,(0.1184)\pm0.0009$ 			& $0.1192\,(0.1191)\pm0.0010$
\tabularnewline
$\bs{\tau_{\mathrm{reio}}}$ 			& $0.0594\,(0.0551)^{+0.068}_{-0.082}$		 & $0.0573\,(0.0630)^{+0.0069}_{-0.0078}$
\tabularnewline
$\bs{\phi_{i}}\,[M_P]$ 				& -- 									& $0.328\,(0.353)^{+0.11}_{-0.062}$	\\
\hline
$H_{0}\, [\mathrm{km/s/Mpc}]$			& $68.21\,(68.18)\pm0.42$ 				& $69.58\,(69.67)\pm 0.80$ 
\tabularnewline
$\Omega_{\rm m}$					& $0.3039\,(0.3048)\pm0.0055$ 			& $0.2940\,(0.2929)\pm0.0071$ 
\tabularnewline
$\sigma_{8}$ 						& $0.8084\,(0.8040)_{+0.0057}^{-0.0063}$ 	& $0.8270\,(0.8310)\pm 0.011$ 
\tabularnewline 
$S_{8}$ 							& $0.814\,(0.810)\pm0.010$	 			& $0.818\,(0.821)\pm0.011$ 
\tabularnewline 
$\Delta G_{N}\,(\%)$					& -- 									& $1.94\,(2.08)\pm 0.94$
\tabularnewline
\hline 
\end{tabular}
\medskip{}
\caption{The mean $\pm1\sigma$ (68\% CL) constraints on the cosmological parameters in $\Lambda$CDM and CCMG, as inferred from the combination of \emph{Planck} 2018 primary CMB data (TT+TE+EE); \emph{Planck} 2018 CMB lensing data; BAO (BOSS DR12) combined with EFT of BOSS and the latest SH0ES $H_0$ constraint. This combination of datasets yields the strongest evidence for the CCMG model, as $\phi_{i}=0.33^{+0.16}_{-0.20}$ $M_{P}$ and $\Delta G_{N}(\%)=1.9\pm 1.8$ with 95\% CL, with $\gtrsim 2\sigma$ significance.
\label{tab:P18+Lens+EFT+BAO+SH0ES}}
\end{table*}

\begin{table}[!htb]
\centering
%\scalebox{0.9}{
$\chi^2$ statistics from the fit to \emph{Planck} 2018 TT+TE+EE +\\
CMB Lensing, BAO + EFT and SH0ES\\
\vspace{2pt}
  \begin{tabular}{|l|c|c|}
    \hline\hline
    			Datasets 									&	LCDM 		&	CCMG	\\
    			\hline \hline
    CMB TT, EE, TE: & &\\
      \;\;\;\; \textit{Planck} 2018 low-$\ell$ TT 					& 22.6			& 22.2		\\
       \;\;\;\; \textit{Planck} 2018 low-$\ell$ EE 					& 396.7 			& 398.3		\\
  \hspace{.16cm} \; \begin{tabular}[t]{@{}c@{}}
  					\textit{Planck} 2018 high-$\ell$ \\ 
  					TT+TE+EE\end{tabular}   				& 2356.0 			& 2355.0		\\

    LSS:& &\\
   \;\;\;\;\,\textit{Planck} CMB lensing							& 9.2 			& 9.1			\\
   \;\;\;\; BAO+EFT (SGC high z)								& 62.7	 		& 64.2		\\
   \;\;\;\; BAO+EFT (NGC low z)								& 70.8 			& 71.0		\\
   \;\;\;\; BAO+EFT (NGC high z)								& 67.1 			& 66.0		\\
   SH0ES 												& 16.6  			& 9.4			\\
    \hline
    Total $\chi^2$   										& 3001.7 			& 2995.2 		\\
    \hline
  \end{tabular}
\medskip{}
  \caption{$\chi^2$ values for the best-fit $\Lambda$CDM and CCMG models, constrained by CMB + CMB Lensing + BAO + EFT + SH0ES. There is reduction of 6.5 in $\chi^2$, for the one additional parameter of the CCMG model.
\label{tab:chi2_P18+Lens+EFT+BAO+SH0ES}}
\end{table}

We also test the CCMG model with another LSS dataset, using effective field theory (EFT) applied to BOSS DR12. this dataset is composed of \emph{Planck} 2018 CMB + lensing + EFT with BAO + SH0ES. The results are tabulated in Table~\ref{tab:P18+Lens+EFT+BAO+SH0ES} (and summarized in Table~\ref{tab:result-summary}). The EFT dataset is less constraining, compared to DES-Y1, as it allows the largest CCMG component of $\phi_{i}=0.33^{+0.16}_{-0.20}$ $M_{P}$ at 95\% CL, which corresponds to a 2\% relative deviation in $\Delta G_{N}$. The CCMG component now raises the Hubble parameter to the value of $H_{0}\!=\!69.6\pm1.6$ km/s/Mpc at 95\% CL, which is the most significant increase in $H_{0}$, compared to the corresponding $\Lambda$CDM benchmark, of all the datasets tested in this work. The Hubble tension reduces, in this scenario, to $2.7\sigma$, compared to $3.9\sigma$ for the $\Lambda$CDM benchmark. Again, the increase comes at the cost of upward shift of $\sigma_{8}$. But the greater reduction in $\Omega_{m}$ results in a mild increase of $S_{8}$, increasing the tension with the combined LSS constraints to just $2.3\sigma$, compared to $2.1\sigma$ for $\Lambda$CDM.

The $\chi^{2}$ statistics, shown in Table \ref{tab:chi2_P18+Lens+EFT+BAO+SH0ES}, indicate a reduction of 6.5 in the total $\chi^{2}$ value, compared to  $\Lambda$CDM. This reduction is once again mainly due to  SH0ES, but we also find a reduction in the  $\chi^{2}$ for some of the LSS likelihoods, resulting in a total increase of only 0.5 due to LSS fits, while the fit to CMB is practically not degraded.

This combination of datasets seems to provide a significant relaxation of the Hubble tension without a substantial damage to the fit to CMB and the LSS data, or a notable increase of the tension between these datasets, represented here by $S_{8}$.

%%%%%%%%%%%%%%%%%%%%%%%%%%%%%%%%%%%%%%%%%%%%%%%%%%%%%%%%%%%%%%%%%%%%%%%%%%%%%
%%%%%%%%%%%%%%%%%%%%%%%%%%%%%%%%%%%%%%%%%%%%%%%%%%%%%%%%%%%%%%%%%%%%%%%%%%%%%
\section{Forecast for CMB-S4 constraints}
%%%%%%%%%%%%%%%%%%%%%%%%%%%%%%%%%%%%%%%%%%%%%%%%%%%%%%%%%%%%%%%%%%%%%%%%%%%%%
%%%%%%%%%%%%%%%%%%%%%%%%%%%%%%%%%%%%%%%%%%%%%%%%%%%%%%%%%%%%%%%%%%%%%%%%%%%%%

In this work we analyzed the CCMG model, which is a special case of a MG model with a scalar field coupled to the Ricci scalar (i.e. fixing $\xi\!=\!-1/6$), because it is symmetric and involves only one additional parameter. Previous analysis of the more general model \cite{Braglia:2020iik} (with $\xi$ free to vary) has found that $\xi=-1/6$ is allowed by constraints from \emph{Planck} 2018, BAO and $H_0$ measurements. Near future experiments may be able to put stronger constraints on the MG model and either exclude or affirm the CCMG model.
Here we consider the planned ground-based CMB 'Stage-4' experiment (CMB-S4).
In order to obtain a forecast for CMB-S4 constraints on the MG model we adopt the expected survey performance of CMB-S4\footnote{CMB-S4 performance expectations: \url{https://cmb-s4.org/wiki/index.php/Survey_Performance_Expectations}} and employ standard Fisher analysis~\cite{Jungman:1995av,Jungman:1995bz,Wu:2014hta}.

The CMB power spectra can be written as
\begin{equation}
C_{\ell}^{XY}=\left(4\pi\right)^{2}\intop dk\,k^{2}\mathcal{T}_{\ell}^{X}\left(k\right)\mathcal{T}_{\ell}^{Y}\left(k\right)P_{\zeta}\left(k\right),
\end{equation}
where $X,Y=\left\{ T,E\right\} $ stand for temperature and E-mode polarization, and $\mathcal{T}_{\ell}^{X}$ are their transfer functions \cite{Seljak:1996is,Yadav:2007ny}. 

\begin{figure}[!htb]
\centering
Forecast on constraints on cosmological parameters in MG scenario from CMB-S4

%\parbox{0.4\textwidth}{
%\begin{tabular}{|l|c|c|}
%\hline\hline
%    Parameter &     best-fit value &   $1\sigma$ \\
%\hline\hline
%	$\xi$ & -0.1667 &      0.0115 \\
%    	$\phi_i$ &  0.297 &      0.005 \\
%\hline
%\end{tabular}
%}
%\,
%\begin{minipage}[c]{0.5\textwidth}
%\centering
\includegraphics[width=.8\columnwidth]{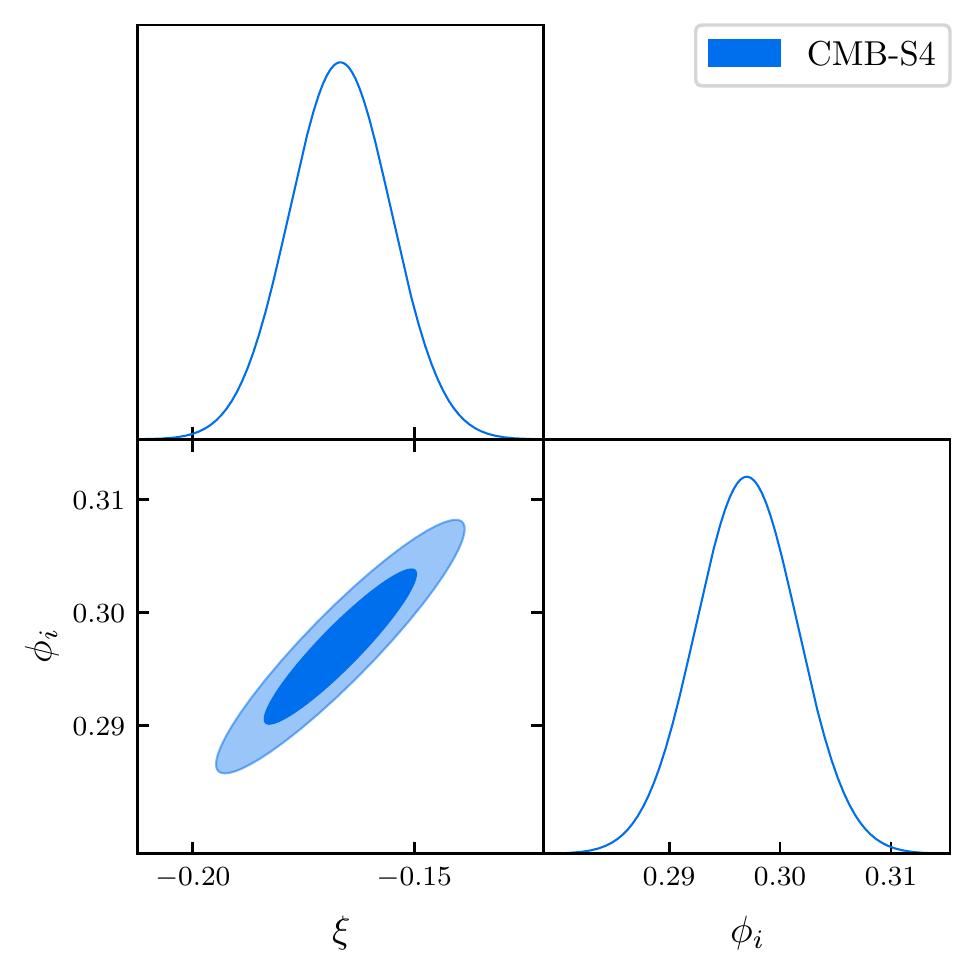}
%\end{minipage}
\caption{Constraints on modified gravity (MG) parameters and $H_0$, expected by the planned CMB-S4 experiment, after maximizing over all the cosmological parameters. We find the constraints $\xi=-1/6\pm 0.0115$ and $\phi_{i}=0.297\pm 0.005$ at 68\% CL, where for $\phi_{i}$ we chose the best-fit value from the analysis of the joint dataset in Table \ref{tab:P18+Lens+BAO+SN+H0}. We can expect CMB-S4 constraints on MG parameters  to be at least an order of magnitude smaller than the values themselves, thus advocating whether the CI scenario is favorable or not. \label{tab:CMB-S4_MG}}
\end{figure}

The forecast on the variance for a set of parameters $\theta_{i}$ may be obtained by defining the Fisher matrix as
\begin{equation}
F_{ij}=\sum_{\ell}\frac{2\ell+1}{2}f_{sky}\text{Tr}\left[C_{\ell}^{-1}\frac{\partial C_{\ell}}{\partial\theta_{i}}C_{\ell}^{-1}\frac{\partial C_{\ell}}{\partial\theta_{j}}\right],
\end{equation}
where $f_{sky}$ is the sky-fraction covered and $C_{\ell}$ are the covariance matrices, which are given by
\begin{equation}
C_{\ell}=\left(\begin{array}{ccc}
\tilde{C}_{\ell}^{TT} & \tilde{C}_{\ell}^{TE} & \tilde{C}_{\ell}^{Td}\\
\tilde{C}_{\ell}^{TE} & \tilde{C}_{\ell}^{EE} & \tilde{C}_{\ell}^{Ed}\\
\tilde{C}_{\ell}^{Td} & \tilde{C}_{\ell}^{Ed} & \tilde{C}_{\ell}^{dd}\\
\end{array}\right),
\end{equation}
where we have defined
\begin{equation}
\tilde{C}_{\ell}^{XY}\equiv C_{\ell}^{XY}+N_{\ell}^{XY},
\end{equation}
where $N_{\ell}^{XY}$ are the noise power spectra, given by
\begin{align}
N_{\ell}^{TT} & =\Delta_{T}^{2}e^{\ell\left(\ell+1\right)\sigma_{b}^{2}}\\
N_{\ell}^{EE} & =2\times N_{\ell}^{TT},\nonumber 
\end{align}
where $\Delta_{T}$ is the temperature sensitivity, $\sigma_{b}=\theta_{\rm FWHM}/\sqrt{8\log2}$, with the full-width-half-maximum $\theta_{\rm FWHM}$ given in radians. For the lensing noise $ N_{ell}^{dd} $ we follow Ref.~\cite{Smith:2010gu}, constructing it from the E and B modes data, then subtracting from the B-mode data which is in turn used again to construct  $ N_{ell}^{dd} $ and so forth, until we reach convergence. We use the 93, 145 and 225 GHz frequencies, with the corresponding sensitivities of $\Delta T=1.5,1.5,4.8 \mu$K-arcmin, resolution of $\theta_{\rm FWHM}=2.2,1.4,1.0$ arcmin, over 40\% of the sky and a prior on the optical depth of reionization of $\tau=0.06\pm0.01$. The CMB-S4 experiment is expected to observe the $\ell$ range between 30 and 5000 for polarization, although the highest modes will be noise-dominated. We also ignore $\ell>3000$ for temperature, as higher multipoles would be contaminated by foregrounds.

Finally we define the correlation matrix as $C_{ij}\equiv F_{ij}^{-1}$, thus the variance of each of the parameters $\Theta_i$ is, according to Cramér–Rao bound, $\sigma_{i}\ge\sqrt{C_{ii}}$. For the fiducial values of the parameters we use the best fit values in Table \ref{tab:P18+Lens+BAO+SN+H0}. The expected constraints on MG parameters from CMB-S4 experiment are shown in Fig. \ref{tab:CMB-S4_MG}. We find that the CMB-S4 experiment is expected to place strong constraints on $\xi$ and can help determine if CI scenario is preferred. We can also expect CMB-S4 to improve constraints from \emph{Planck} on the evidence for physics beyond $\Lambda$CDM (e.g. EDE, CCMG).

%%%%%%%%%%%%%%%%%%%%%%%%%%%%%%%%%%%%%%%%%%%%%%%%%%%%%%%%%%%%%%%%%%%%%%%%%%%%%
%%%%%%%%%%%%%%%%%%%%%%%%%%%%%%%%%%%%%%%%%%%%%%%%%%%%%%%%%%%%%%%%%%%%%%%%%%%%%
\section{Discussion and Conclusions}
%%%%%%%%%%%%%%%%%%%%%%%%%%%%%%%%%%%%%%%%%%%%%%%%%%%%%%%%%%%%%%%%%%%%%%%%%%%%%
%%%%%%%%%%%%%%%%%%%%%%%%%%%%%%%%%%%%%%%%%%%%%%%%%%%%%%%%%%%%%%%%%%%%%%%%%%%%%

\begin{table*}[ht!]
\centering
Constraints summary on CCMG for varying data sets  \\
\vspace{2pt}
  \begin{tabular}{|l|c|c|c|c|}
    \hline \hspace{0.2cm} Parameter & \begin{tabular}[t]{@{}c@{}}\emph{Planck}  2018 \\ TT+TE+EE\end{tabular} & \begin{tabular}[t]{@{}c@{}}\emph{Planck} 2018\\ TT+TE+EE, \\  CMB lensing,  BAO,\\ RSD, SNIa\\ and SH0ES\end{tabular} & \begin{tabular}[t]{@{}c@{}}\emph{Planck} 2018\\ TT+TE+EE, \\  CMB lensing,  BAO,\\ RSD, SNIa,\\ SH0ES \\ and DES-Y1\end{tabular}& \begin{tabular}[t]{@{}c@{}}\emph{Planck} 2018\\ TT+TE+EE, \\  CMB lensing,\\ BAO+EFT \\ and SH0ES\\ \end{tabular}\\ 
\hline
\hline
$\bs{\phi_i}\,[M_P]$					& $< 0.213 $							& $ 0.297^{+0.11}_{-0.075} $
										& $ 0.264^{+0.13}_{-0.072} $				& $ 0.328^{+0.11}_{-0.062} $\\
\hline
$\Delta G_N\,(\%)$ 						& $0.68^{+0.14}_{-0.75}$ 					&$1.70^{+0.81}_{-1.1}$ 
										&$1.36^{+0.68}_{-1.1}$					& $ 1.94\pm0.94 $\\

$H_0 \, [\mathrm{km/s/Mpc}]$ 				& $67.98^{+0.63}_{-1.1}$ 					& $69.24^{+0.60}_{-0.83}$
										& $69.40^{+0.60}_{-0.75}$ 				& $ 69.58\pm0.80 $\\

$\sigma_8$ 							& $0.8170^{+0.022}_{-0.020}$ 				& $0.8242^{+0.0090}_{-0.012}$ 
										& $0.8176^{+0.0087}_{-0.0110}$ 			& $ 0.8270\pm0.011 $\\
										
$S_8$	 							& $0.830\pm0.016$ 						& $ 0.820\pm0.010$ 
										& $0.809\pm0.009$ 					& $ 0.818\pm0.011 $\\
\hline
$\Delta\chi^2$							& $+4.8$ 						& $-9$ 		
										& $-0.8$				 		& $-6.5$\\
\hline
 \end{tabular}
\medskip{}
\caption{
The mean $\pm1\sigma$ constraints on cosmological parameters in the CCMG scenario from \emph{Planck} 2018; CMB lensing; BAO; SNIa; SH0ES; RSD; DES-Y1; and a combined BAO+EFT dataset. Only $\phi_i$ is a sampled parameter. The significance of the CCMG component is highly dependent on the datasets: the inclusion of SH0ES tends to increase the value of $\phi_{i}$, whereas the inclusion of DES-Y1 reduces its value. The right column refers to another dataset composed of the BAO+EFT likelihood, which allows for a larger value for $\phi_{i}$. Even the highest value found for $H_{0}$ does not relieve the Hubble tension completely.
\label{tab:result-summary}}
\end{table*}

\begin{figure*}[!htb]
\centering
Constraints on CCMG parameter from varying sets of data
\vspace{2pt}
\includegraphics[width=2\columnwidth]{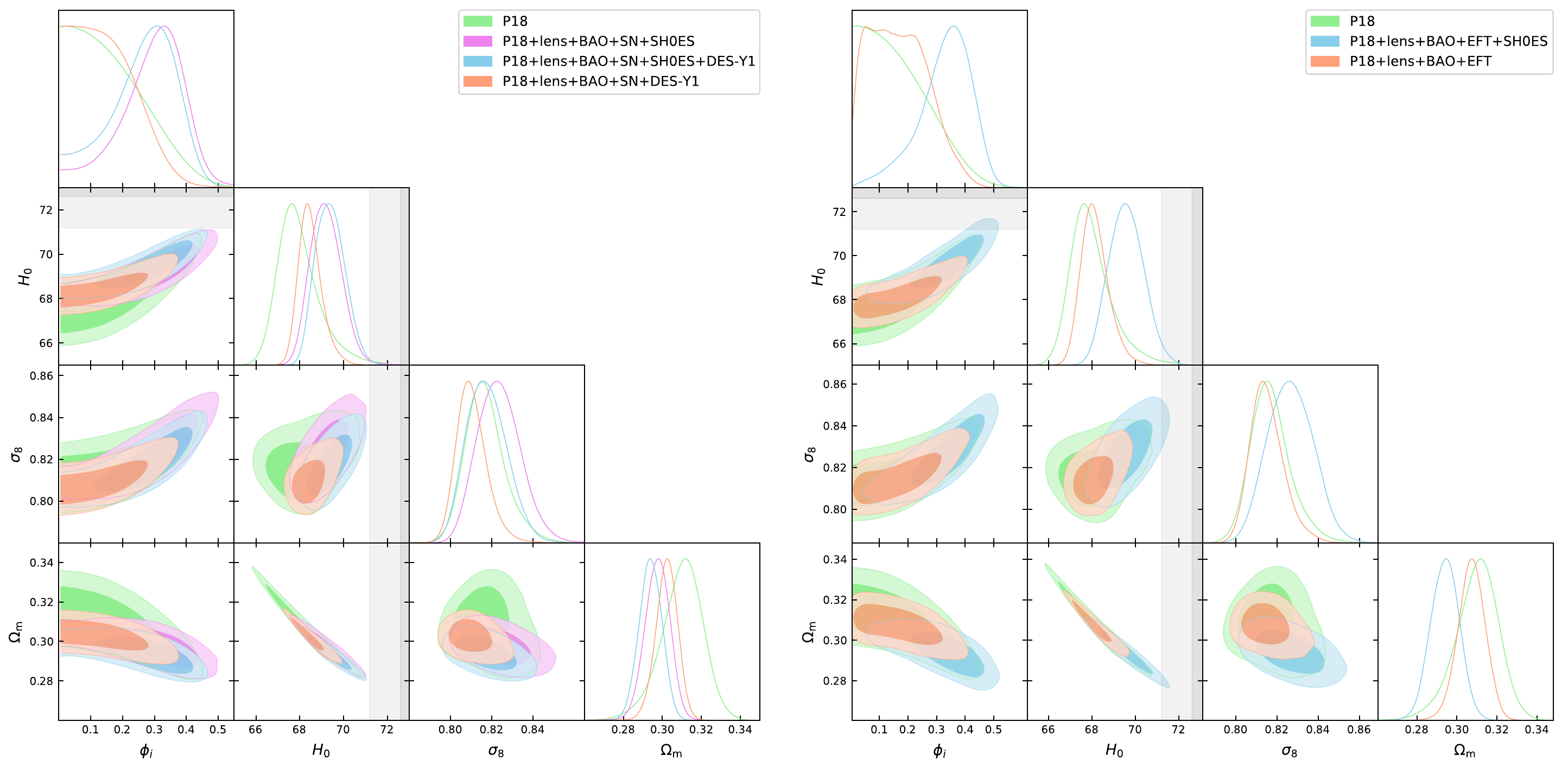}
\caption{Constraints on the CCMG parameter from various datasets: primary \emph{Planck} 2018; CMB lensing; BAO; RSD; SNIa; SH0ES; DES-Y1; and a combined BAO+EFT. Here we present a subset of the parameters: the initial condition $\phi_i$ for the CCMG, along with $H_0$ [km/s/Mpc] and $\sigma_8$. The contours show $1\sigma$ and $2\sigma$ posteriors for various dataset combinations, computed with GetDist \cite{Lewis:2019xzd}. The P18 dataset alone ({\it green}) yields a posterior for $\phi_{i}$ which tends to zero, thus disfavoring a significant CCMG component, in contrast to the combined datasets ({\it purple} and {\it blue}), which feature a significant CCMG component. The EFT dataset seems to put lower constraints on $\sigma_{8}$ than DES-Y1, as shown in its posteriors for each dataset, which explains the more significant CCMG component when using the EFT instead of DES-Y1 ({\it blue} on each side). Here we also include datasets without SH0ES ({\it salmon}), in which the CCMG significance is almost completely erased, indicating the low preference of the model by all other datasets.
\label{fig:CCMG-summary}}
\end{figure*}

In recent years the discrepancy between the values of the Hubble constant $H_0$, the current expansion rate of the Universe, inferred from early-Universe measurements, such as \emph{Planck} 2018 CMB, and late-Universe measurements, such as the SH0ES collaboration distance-ladder measurements, has reached $\gtrsim 4\sigma$ confidence. However, other independent late-Universe experiments report lower values of $H_{0}$, in better agreement with early-Universe inferred values, such as the TRGB-based calibration \cite{Freedman:2020dne} of the local distance-ladder, which yields $H_{0}=69.6\pm0.8$ km/s/Mpc, as well as the recent TDCOSMO+SLACS analysis \cite{Birrer:2020tax}, which reported $H_{0}=67.4^{+4.1}_{-3.2}$ km/s/Mpc.

A recent review, Ref.~\cite{Knox:2019rjx}, of the phenomenology of the Hubble tension suggests that, to restore concordance between recent cosmological data and the cosmological model, increasing the value of $H_{0}$ alone is not enough, but one should reduce the value of the sound horizon at last scattering $r_{s}^{\star}$ as well. It was also suggested that the most promising method to accomplish this goal is by introducing new physics just prior to recombination, at the proximity of matter-radiation equality, which would trigger a rapid increase of the expansion rate throughout this period. A typical way of realizing this methodology is to introduce a new energy component to the cosmological model, so that it will increase $H(z)$ throughout this period and then dilute fast enough to be negligible at later epochs.

In this work we considered the CCMG model as a candidate for alleviating the Hubble tension. We analyzed it using various combinations of datasets composed of early-Universe data, direct measurements of $H_{0}$ and LSS data: \emph{Planck} 2018 CMB and its lensing, BAO (6dF, SDSS DR7 and SDSS DR12), SDSS DR12 RSD, SN distance data from Pantheon, SH0ES distance-ladder measurements of $H_{0}$, DES-Y1 3x2pt and BAO+EFT (BOSS/SDSS galaxies clustering DR12). We also compared the results to a similar analysis done for the EDE model~\cite{Hill:2020osr}. The constraints we found on the CCMG model and its influence on other cosmological parameters share many of the characteristics of the EDE model, while the former introduces only one additional parameter to the cosmological model and is not fine tuned as the latterl, which requires at least two additional parameters. We find that the CCMG model allows an increase of the Hubble parameter up to $H_{0}=69.58\pm0.80$ km/s/Mpc, when considering all types of datasets (summarized results in Table~\ref{tab:result-summary} and Fig.~\ref{fig:CCMG-summary}).

Initially we considered primary CMB anisotropies alone: \emph{Planck} 2018 TT+TE+EE. Although the value of $H_0$ is increased slightly, there is no significant evidence for the CCMG model (Table~\ref{tab:CCMG-P18}). Furthermore, the total fit to the CMB is worsened in the CCMG scenario, compared to  $\Lambda$CDM. We conclude that the CCMG model is not preferred by primary CMB data alone. 
In contrast to the EDE scenario, in which the posterior of the EDE component might be biased due to degeneracy of the other parameters of the model (as describe in Ref.~\cite{Smith:2020rxx}), the posterior of $\phi_{i}$, shown in Fig.~\ref{fig:CCMG-summary},  indicates accurately the preference of the dataset. For the P18 dataset it is located around zero, indicating it disfavours CCMG.

When we supplement the primary CMB dataset with \emph{Planck} 2018 lensing + BAO + RSD + SNIa + SH0ES, we find, as shown in Table~\ref{tab:P18+Lens+BAO+SN+H0}, a substantial CCMG component, corresponding to $\phi_{i}=0.3^{+0.14}_{-0.24}$ $M_{P}$ and $H_{0}=69.24\pm1.4$ km/s/Mpc at \%95 CL. The tension with the SH0ES measurements is reduced to $3.1\sigma$, while the tension with LSS data, $2.4\sigma$, is only slightly greater than that of the $\Lambda$CDM benchmark. The CCMG model offers a better fit to this combined dataset as $\Delta\chi^{2}=-9$. This reduction of $\Delta\chi^{2}$ is mainly due to the better fit to SH0ES data, as shown in Table~\ref{tab:chi2_P18+Lens+BAO+SN+H0}. We note that the better fit to both CMB and SH0ES data comes at the expense of a worse fit to BAO and lensing data (the LSS part of this dataset), indicating a correlation between the different datasets.
As described in Ref.~\cite{Hill:2020osr} for the EDE model, the introduction of the CCMG component forces some cosmological parameters to shift in order to keep the fit to the CMB data. But, due to the increase in the gravitational strength and the non-localized dynamics of the CCMG model, the shifts in $\Omega_{c}h^{2}$ and $n_{s}$ are very small compared to the EDE model. However, for the same reason, the downward shift in $\Omega_{m}$ is more significant, suppressing the increase in $S_{8}$ due to the increase in the matter clustering amplitude $\sigma_{8}$. The correlation between the cosmological parameters is shown in Fig.~\ref{fig:P18+Lens+BAO+SN+H0}.

Including  the DES-Y1 likelihood in the combined dataset, we saw that the  posterior of $\phi_{i}$ is driven slightly backwards (Fig.~\ref{fig:CCMG-summary}), corresponding to a smaller CCMG component. The inclusion of DES-Y1 likelihood acts to reduce the value of $S_{8}$, due to the stronger constraints on $\Omega_{m}$, resulting in smaller tension with the LSS data. Nevertheless, the fit to LSS datasets is worse than that of the $\Lambda$CDM benchmark, as the fit to BAO+RSD (BOSS DR12) dataset worsens compared to the improvement exhibited by the benchmark, as shown in Table~\ref{tab:chi2_P18+Lens+BAO+SN+H0+DES_Y1}. Excluding SH0ES from the combined dataset erases the CCMG component, as shown in Fig.~\ref{fig:CCMG-summary}. These results affirm the correlation between the datasets, which leads to the conclusion that it is not possible to reconcile DES, BAO and SH0ES datasets simultaneously, using the methodology of reducing $r_{s}^{\star}$, as recently elaborated  in Ref.~\cite{Jedamzik:2020zmd}.

We also considered a combined dataset which includes the newly published BAO+EFT likelihood. We found a larger CCMG component than in any other combination of datasets,  $\phi_{i}=0.33^{+0.16}_{-0.20}$ $M_{P}$ at \%95 CL, which corresponds to $H_{0}=69\pm1.6$ km/s/Mpc (which is close to the most recent values from the TRGB and TDCOSMO analyses). The relatively large CCMG component is followed by a large value for $\sigma_{8}$, but due to the significant reduction in $\Omega_{m}$ (Table~\ref{tab:P18+Lens+EFT+BAO+SH0ES}), which is allowed by this dataset, the resulting value of $S_{8}$ is not increased as one would have expected (compared to the values in Table~\ref{tab:P18+Lens+BAO+SN+H0}, for example). Excluding SH0ES from this dataset also results in significant decrease of the CCMG component, however not to the same extent as in the datasets with DES, as shown in Fig.~\ref{fig:CCMG-summary}.

Finally, our forecast for the near-future ground experiment CMB-S4 shows that we can expect it to place strong  constraints on the parameter $\xi$ to distinguish the CCMG model from the more general MG, and also to place stronger constraints on $\phi_{i}$. In light of the results of our analysis, the CCMG model might present a rather elegant and natural solution to relieve the Hubble tension, compared to  EDE, but since no model that acts to increase $H_{0}$ by reducing $r_{s}^{\star}$ seems to be able to reconcile BAO, DES and SH0ES at the same time, the search for new physics to explain the Hubble tension has not concluded. 

\medskip
\noindent\paragraph*{Note added:} While this paper was undergoing the last round of text edits, Ref.~\cite{braglia2020early} appeared on the arXiv with overall consistent conclusions in its short discussion of the model analyzed in depth here.  

\acknowledgements
We thank Sunny Itzhaki and Marc Kamionkowski for useful discussions and Tristan Smith, Mikhail Ivanov, Miguel Zumalacarregui and especially Colin Hill and Jos\'e Luis Bernal for tremendous help with the modified CMB codes and the implementation of the various likelihoods in our MCMC analyses. We thank the anonymous referee for helpful suggestions. EDK is supported by a faculty fellowship from the Azrieli Foundation.

\bibliography{References}

\end{document}